\documentclass[prb,twocolumn,showpacs,preprintnumbers,amsmath,amssymb]{revtex4}

\usepackage{graphicx}
\usepackage{dcolumn}
\usepackage{bm}
\usepackage{textcomp}

\begin{document}


\title{Structural phase transitions in epitaxial perovskite films}

\author{Feizhou He}
 \email{fhe@phys.uconn.edu}
\author{B. O. Wells}%
 \affiliation{Department of Physics, University of Connecticut,
Storrs, Connecticut 06269}%

\author{Z. -G. Ban}
\author{S. P. Alpay}%
 \affiliation{Department of Materials Science and Engineering and Institute of Materials Science, University of Connecticut,
Storrs, Connecticut 06269}%

\author{S. Grenier}
\author{S. M. Shapiro}
\author{Weidong Si}
\affiliation{Department of Physics, Brookhaven National Lab,
Upton, New York 11973}%

\author{A. Clark}
\author{X. X. Xi}
\affiliation{Department of Physics, The Pennsylvania State
University, University Park, Pennsylvania 16802}

\date{\today}

\begin{abstract}
Three different film systems have been systematically investigated
to understand the effects of strain and substrate constraint on
the phase transitions of perovskite films. In SrTiO$_3$ films, the
phase transition temperature T$_C$ was determined by monitoring
the superlattice peaks associated with rotations of TiO$_6$
octahedra. It is found that T$_C$ depends on both SrTiO$_3$ film
thickness and SrRuO$_3$ buffer layer thickness. However, lattice
parameter measurements showed no sign of the phase transitions,
indicating that the tetragonality of the SrTiO$_3$ unit cells was
no longer a good order parameter. This signals a change in the
nature of this phase transition, the internal degree of freedom is
decoupled from the external degree of freedom. The phase
transitions occur even without lattice relaxation through domain
formation. In NdNiO$_3$ thin films, it is found that the in-plane
lattice parameters were clamped by the substrate, while
out-of-plane lattice constant varied to accommodate the volume
change across the phase transition. This shows that substrate
constraint is an important parameter for epitaxial film systems,
and is responsible for the suppression of external structural
change in SrTiO$_3$ and NdNiO$_3$ films. However, in SrRuO$_3$
films we observed domain formation at elevated temperature through
x-ray reciprocal space mapping. This indicated that internal
strain energy within films also played an important role, and may
dominate in some film systems. The final strain states within
epitaxial films were the result of competition between multiple
mechanisms and may not be described by a single parameter.
\end{abstract}

\pacs{68.18.Jk, 68.55.Jk, 77.55.+f, 77.80.-e}
\maketitle

\section{Introduction}

Perovskite thin films have received great interest in recent years
due to a variety of interesting properties such as high-T$_C$
superconductivity, colossal magnetoresistivity (CMR),
ferroelectricity and metal-insulator transition. While bulk
crystals of perovskites have been well studied, thin films are not
understood in nearly the same details. It is well known that the
properties of thin films can be quite different from corresponding
bulk materials. The reason is often believed to be the strain and
high concentration of defects in the thin films. This work
illustrates that another important parameter for films is the
geometrical constraint imposed by the substrates. Due to the
excellent epitaxy, the films are tied to the underlying substrate,
therefore the in-plane lattice parameters of films are not free to
attain their bulk equilibrium values. The macroscopic sizes of
films are always the same as that of substrates  This is known as
the substrate clamping effect, or substrate constraint, which
forces the temperature dependence of in-plane lattice parameters
of films to follow that of substrates. This effect is often
neglected but as discussed in this paper, it has great effect on
phase transitions and domain formation in films. The strain and
substrate constraint give us a method to fine-tune the properties
of the thin films, or even create a new phase that is not possible
in bulk form. In recent years there are emerging theoretical
efforts to treat the effects of
strain,\cite{Pertsev00,Pertsev98,Roytburd97a,Ban02} based on
thermodynamic analysis using Landau theory of phase transition.

SrTiO$_3$ (STO) has long been an important model system for
condensed matter physics. More recently it has received attention
for having a large and variable dielectric constant, making it
ideal for tunable microwave devices.\cite{Xi00} Bulk STO crystals
are cubic perovskite at room temperature, with space group
Pm$\overline{3}$m (O$^{1}_{h}$), but become tetragonal, space
group I4/mcm (D$^{18}_{4h}$), below 105K. This phase transition
was extensively studied in the 1960's and 1970's as the
prototypical example of a soft-mode phase
transition,\cite{Shirane69} involving a progressive softening of
the zone boundary phonon mode due to the TiO$_6$ octahedra
rotation about a formerly cubic axis. When the energy for this
mode at the zone boundary reaches zero, the lower symmetry
structure condenses. Since the phonon condensation occurs at the
zone boundary, this phase transition is non-polar and
antiferrodistortive. The high temperature structure of SrTiO$_3$
is the same as that of BaTiO$_3$ (BTO), the canonical
ferroelectric crystal. In BTO, the ferroelectric phase transition
is associated with a softening of a zone-center phonon mode in
which the central Ti ion shifts with respect to the oxygen cage.
For SrTiO$_3$, the ferroelectric phase, which almost occurs at
lower temperatures, is suppressed by quantum
fluctuations.\cite{Muller79} However the ferroelectric state can
be induced under pressure,\cite{Uwe76} applied electric
fields,\cite{Hemberger96} by doping impurities,\cite{Bednorz84}
isotope exchange,\cite{Itoh99} or in thin films,\cite{Fuchs99}
indicating that this phase transition is sensitive to lattice
strain. Therefore, studying the properties of perovskite films
under strain conditions may as well help to better understand the
ferroelectric phase transitions in related systems, such as
BaTiO$_3$ and PbTiO$_3$. In fact, similar structural phase
transitions occur in most perovskite-based oxide crystals, for
example SrRuO$_3$. Therefore, the results we obtained here may be
generalized to other perovskite films systems.

SrRuO$_3$ (SRO) is a ferromagnetic perovskite with Curie
temperature of about 160K. It has excellent electrical
conductivity and is chemically stable. Since SRO has good lattice
matching with other perovskites, it is the material of choice for
electrodes of perovskite-based devices. However, a bottom SRO
layer may change the strain state and thus the properties of the
top layer. A detailed study on effects of buffer layers is
necessary, so that we can either minimize the negative influence,
or utilize the effect to tune the strain state of the layers
above. Here we investigated the influence of SRO buffer layers on
the phase transition of STO films, and the domain structures of
SRO layers.

We also studied the behavior of NdNiO$_3$ (NNO) thin films under
biaxial strain. Perovskite nickelates (RNiO$_3$ with R~=~rare
earth) show metal-insulator (MI) transitions with the transition
temperature rising systematically as the size of the rare earth
decreases, which implies an increase in the distortion away from
the ideal cubic perovskite. This transition is first-order in
nature and associated with volume change across the phase
transition. Bulk NNO has shown large pressure dependence of the MI
transition,\cite{Canfield93} with change in T$_C$ of over 100K
under high pressure. Similar effects have been observed in
epitaxial NNO films,\cite{Tiwari02} but the strain state across
the MI transition and the microstructure of NNO films are not
clear.

In earlier work, we reported experimental results of the strain
effects on the phase transitions of STO thin films, and the
observations of the smooth evolution of the out-of-plane lattice
parameters across the transition.\cite{He03} A theoretical
analysis has also been undertaken to understand the strain
relaxation mechanism in epitaxial STO films and the influence of
the SRO buffer layers.\cite{Ban04} In this paper, a systematic
study of the misfit strain and substrate clamping effects on phase
transitions of perovskite films is reported. We show that the
thickness of both thin films and buffer layers may influence
T$_C$, and the temperature dependence of the lattice relaxation is
determined by the substrate clamping effect and the volume change
associated with the phase transition. The mechanisms involved in
determining the final strain states in films are discussed.

\section{Experiment}

The STO films were grown by pulsed laser deposition (PLD) on
LaAlO$_3$ (LAO) single crystal substrates. Before growing the STO
films, SrRuO$_3$ buffer layers of various thickness were
deposited. The SRO buffer serves as both a method to tune the
strain state in the STO films and as a base electrode for
electrical measurements. The samples are of the STO/SRO/LAO type,
with STO film thickness range from 10 nm to 1 $\mu$m and SRO
buffer layers from 0 nm to 350 nm. Details of the sample
preparation are available elsewhere.\cite{Li98a} We report on
experiments of both the STO and SRO layers.

The NdNiO$_3$ thin films, with thickness of 60 nm and 200 nm, were
grown on LAO substrates by PLD. A KrF excimer laser was used with
an energy density of $\sim$2.0 J/cm$^2$ and a repetition rate of
5~Hz. The substrate was heated to 700\textcelsius\ and an oxygen
pressure of 200 mTorr was used during the deposition. After
deposition, the films were cooled to room temperature at a rate of
60\textcelsius\ per minute. Resistivity was measured by the
standard four probe method through 5~K to 300~K with current about
0.5 $\mu$A.

X-ray diffraction (XRD) measurements were carried out at beamline
X22A and X22C at the National Synchrotron Light Source, Brookhaven
National Laboratory. Synchrotron radiation has the advantages of
high intensity and small divergence angle, which are critical for
studying very thin films grown on substrates and buffers of
similar structures. X22A has a bent Si(111) monochromator, giving
a small beam spot and a fixed incident photon energy of 10 keV.
The longitudinal resolution with a Si(111) analyzer was at least
0.001 \AA$^{-1}$ (HWHM) for an (002) peak, as measured from the
substrate. X22C has a Ge(111) double monochromator, focused in
vertical plane, with variable energy range 3$\sim$12 keV.  The
samples ware cooled in a closed cycle refrigerator with a
temperature control better than $\pm$0.5 K. Above room temperature
the sample was heated in a high temperature capable displex with a
base temperature near 10 K and a maximum of 800 K.

Throughout this paper we list most peaks with reference to a cubic
cell, unless otherwise specified. For cubic notations, we define
the axis normal to the surface of the substrate is the $c$ axis.
Thus our scattering peaks, and reciprocal lattice positions, are
listed as ($h$ $k$ $l$) with $l$ the reciprocal direction
perpendicular to the surface. For tetragonal or orthorhombic
phases, a subscript $t$ or $o$ is added to the peak index, and the
orientation is redefined accordingly.

\section{Results}

\subsection{SrTiO$_3$}

\begin{figure}
\includegraphics[scale=1.0]{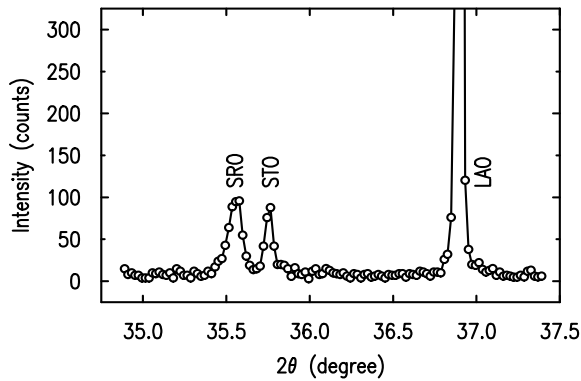}
\caption{\label{fig:002peak}Raw XRD scan for (0~0~2) peaks on 200
nm STO / 350 nm SRO / LAO sample, showing well resolved peaks.
Photon energy is 10 keV.}
\end{figure}

The high intensity and the collimation of the synchrotron x-ray
source allows for scattering from very thin films in a high
resolution mode to easily separate film-buffer-substrate peaks. A
typical scan over the (0~0~2) peaks is shown in
Fig.~\ref{fig:002peak}. The STO and SRO peaks are well resolved,
moreover the bulk LAO peak can be used as an internal reference to
minimize systematic error. X-ray diffraction and TEM (not shown
here) show that all the thin films are in excellent epitaxy with
the LAO substrates and SRO buffer layers.

\subsubsection{The Phase Transition}

In bulk STO, the soft mode phase transition occurs at 105 K. This
structural phase transition from cubic to tetragonal symmetry
involves the rotation of the TiO$_6$ octahedra around [0~0~1]
axis. Because the TiO$_6$ octahedra share the corner oxygen atoms,
the neighboring TiO$_6$ rotate in opposite directions. This leads
to a tetragonal lattice with space group I4/mcm. The unit cell of
the tetragonal phase is about $\sqrt{2}a\times\sqrt{2}a\times2a$,
where $a$ is the lattice parameter of original cubic unit cell,
therefore the volume of the tetragonal unit cell is about four
times as that of the cubic unit cell. This tetragonal phase is a
superlattice of the original cubic phase. The selection rule of
the diffraction peaks is (in
tetragonal notation): \\
$hkl:h+k+l=2n;~~~0kl:l=2n;~~~h0l:l=2n;$\\
$hhl:l=2n;~~~h\bar{h}l:l=2n$\\
In pseudo-cubic notation, the additional superlattice peaks are at
half integer positions, such as (1/2, 1/2, 3/2)$_c$.

\begin{figure}
\includegraphics[scale=1.0]{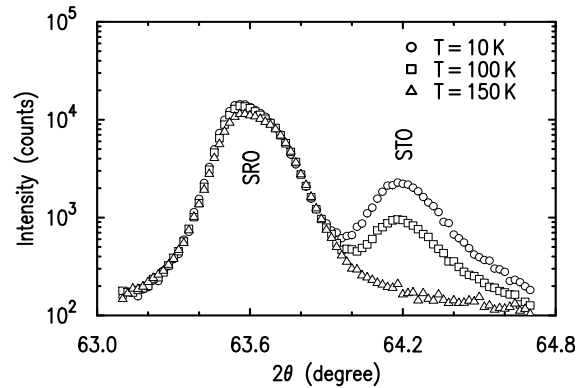}
\caption{\label{fig:sto-super}Temperature dependence of (1/2, 1/2,
7/2) STO superlattice peak for 1 $\mu$m STO / 350 nm SRO / LAO
sample. The other peak is from SRO layer.}
\end{figure}

\begin{figure}
\includegraphics[scale=1.0]{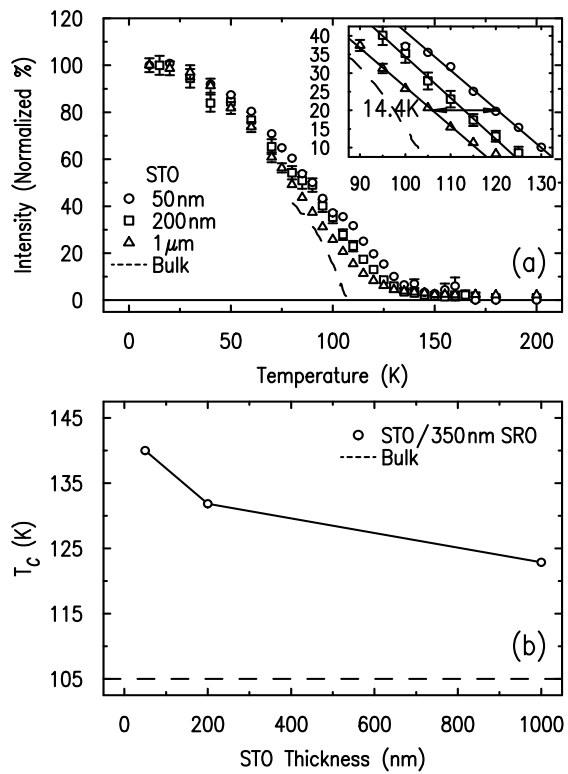}
\caption{\label{fig:sto-tc}Temperature dependence of superlattice
peak intensity on STO film thickness. Bulk value from
Ref.~\onlinecite{Hayward99}.}
\end{figure}

In STO films, we also observed the appearance of the superlattice
peaks below T$_C$, as shown in Fig.~\ref{fig:sto-super}, which
indicates that the phase transition also occurs in thin films. The
temperature dependence of the peak intensity, as seen in
Fig.~\ref{fig:sto-tc}, shows obvious rounded tails around T$_C$,
which means the phase transitions occur over a range of
temperature. This is considered normal for films due to strain and
defects, but it makes determination of T$_C$ more difficult. We
chose to linearly extrapolate the part of curves between 10\% and
40\% of maximum intensity, and define the T$_C$ as where this line
crosses the 0\% line. This method is justified by both our data as
well as in literature.\cite{Doi00} Though it might give a T$_C$
slightly higher than the true transition, we believe the
comparison between different films is accurate.

Our first set of samples have STO film thickness varying from 50
nm to 1 $\mu$m, with the same SRO buffer layer thickness of 350
nm. This shows us how the phase transition depends on the
thickness of STO films, as other conditions keep the same. The
T$_C$ difference, $\Delta$T$_C$, is about 14 K between the 50 nm
sample and 1000 nm sample, as seen in Fig.~\ref{fig:sto-tc}. The
thinner film has higher T$_C$, while thicker film has T$_C$ closer
to the bulk value. The in-plane tensile strain measured by XRD at
room temperature increases from 0.01\% in 1000 nm film to 0.23\%
in 50 nm film. Therefore in this case, T$_C$ increases
monotonically with increasing in-plane strain.

\begin{figure}
\includegraphics[scale=1.0]{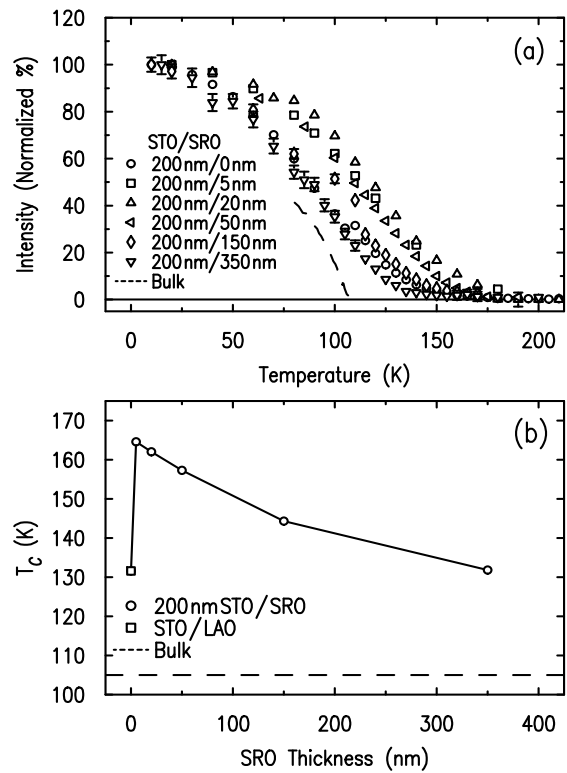}
\caption{\label{fig:sro-tc}Temperature dependence of superlattice
peak intensity on samples with different SRO thickness. Bulk value
from Ref.~\onlinecite{Hayward99}.}
\end{figure}

The second set of samples have the same STO thickness of 200 nm,
but SRO thickness ranges from 0 nm to 350 nm. This gives us
information about the influence of the SRO buffer layers. We found
that varying the SRO thickness has even bigger effect on T$_C$, as
shown in Fig~\ref{fig:sro-tc}. The thinnest (5~nm) SRO layer
causes the highest T$_C$ in STO films, which is about 50 K above
the bulk value. When SRO thickness increases, T$_C$ decreases.
However, room temperature XRD measurement shows that the in-plane
strains in these samples are still quite small, with the largest
one only about 0.25\%, so this amount of change in T$_C$ is fairly
remarkable. It is interesting to notice that with SRO thickness
decreased to 0 nm, i.e., STO films grown directly on LAO
substrate, the T$_C$ goes back to about 130 K. This signals that
the SRO layers have a significant effect on the strain states of
STO films.

\subsubsection{Lattice Parameters}

The lattice constants were measured by conventional diffraction as
well as grazing incidence X-ray diffraction (GIXD) technique. Due
to the high intensity of the synchrotron radiation, peaks from all
three layers, even as thin as 20 nm, can be seen clearly.
Measuring multiple peaks in (0 0 $l$) direction gives a very
accurate measurement of the $c$ lattice constant. In-plane lattice
constants are obtained either by measuring a partially in-plane
peak such as (2 0 2)/(0 2 2) and then triangulating using the (0 0
$l$) values, or by measuring a (2 0 0)/(0 2 0) type reflection
using grazing incidence diffraction.

\begin{figure}
\includegraphics[scale=1.0]{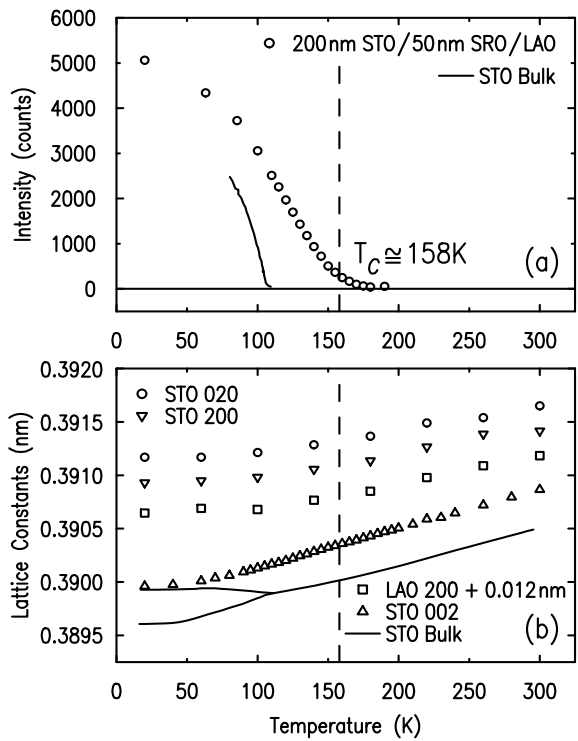}
\caption{\label{fig:sto-lat}Temperature dependence of superlattice
peak intensity, in-plane and out-of-plane lattice parameters of
STO films. Dashed lines indicate T$_C$. LAO and STO bulk data
\cite{Okazaki73} are shown for comparison.}
\end{figure}

We have measured the temperature dependence of both out-of-plane
$c$ of all samples and in-plane lattice parameters of some
samples. Fig.~\ref{fig:sto-lat} shows the comparison between the
lattice constants of films and STO bulk, as well as the LAO
substrate. We believe the difference in (2 0 0) and (0 2 0) is due
to experimental error. It is immediately clear that the lattice
parameters evolve smoothly from low temperature up to room
temperature, without showing any indication of the structural
phase transition. The in-plane lattice constants follow that of
the LAO substrate, which implies that the substrate plays the
controlling role here.

We attribute the smooth variation of the lattice parameters to the
constraint applied by the epitaxy of the film on the far thicker
substrate. This substrate clamping effect forces the in-plane
lattice parameters of films to follow that of substrates.
Therefore, even above the phase transition temperature, as defined
by intensity of superlattice peaks, STO still has different
in-plane and out-of plane lattice constants. In fact, the symmetry
of the films at high temperature is no longer cubic, but
high-symmetry tetragonal, possibly P4/mmm. In other words, now the
phase transition is no longer from cubic Pm$\overline{3}$m to
tetragonal I4/mcm, but from high symmetry tetragonal to lower
symmetry tetragonal. In bulk STO, the splitting of the lattice
parameters at T$_C$ results in domain formation and thus lattice
relaxation. However, in films, there is no change in the shape of
the unit cell during the transition. Therefore, externally, the
films always appear as in single domain, and the phase transitions
occur without lattice relaxation. Obviously the distorted unit
cell favors the TiO$_6$ rotation, so the starting point of the
rotation, T$_C$, is much higher in films. Origins of this effect
are discussed more fully in Section \ref{sec:Discussion}.

\subsection{NdNiO$_3$}

Bulk NdNiO$_3$ is different from bulk SrTiO$_3$ in that it shows
discontinuity in temperature dependence of unit cell volume and
lattice parameters across the first-order MI
transition.\cite{GM92} To understand how they behave in films
under the influence of strong substrate clamping effect, we have
measured the temperature dependence of in-plane and out-of-plane
lattice parameters, as well as resistivity, of NNO films.

\begin{figure}
\includegraphics[scale=1.0]{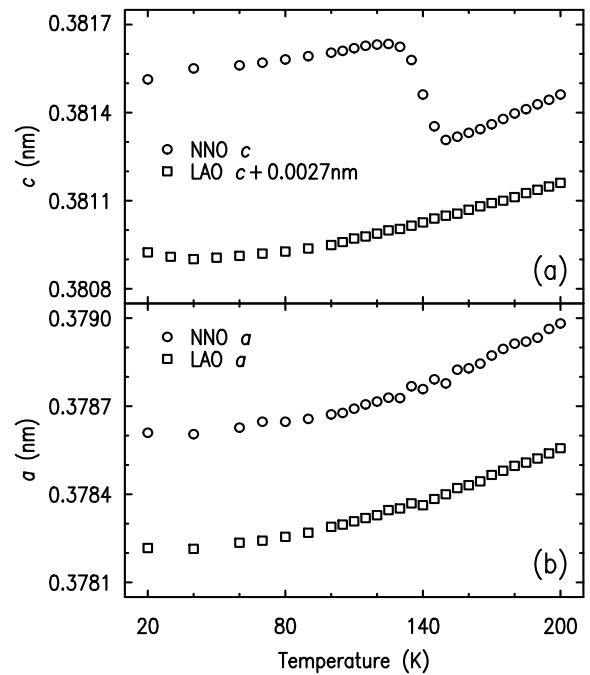}
\caption{\label{fig:nno-lat}Out-of-plane and in-plane lattice
parameters of NdNiO$_3$ films and LAO substrates during heating
up, shown in pseudo-cubic notation.}
\end{figure}

\begin{figure}
\includegraphics[scale=1.0]{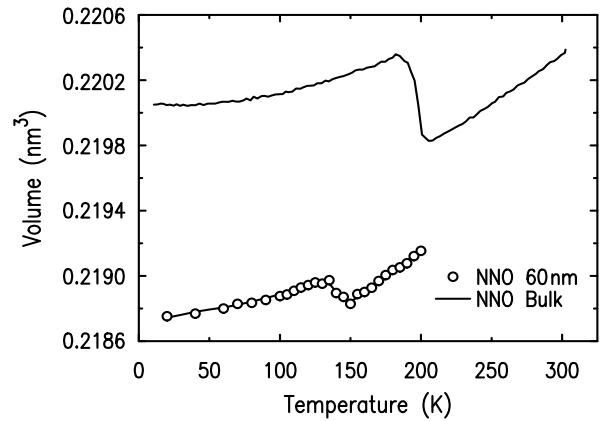}
\caption{\label{fig:nno-v}Temperature dependence of NdNiO$_3$ unit
cell volume. Bulk data from Ref.~\onlinecite{GM92}.}
\end{figure}

\begin{figure}
\includegraphics[scale=1.0]{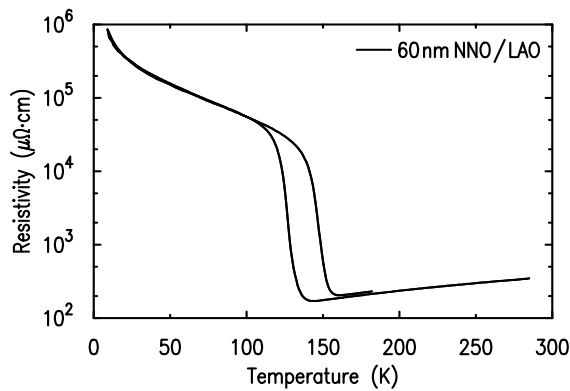}
\caption{\label{fig:nno-r}Resistivity of NdNiO$_3$ films.}
\end{figure}

As shown in Fig.~\ref{fig:nno-lat} for the 60 nm sample, the
in-plane lattice parameters show the same behavior as in STO
films, following closely with the LAO substrates. But the
out-of-plane lattice parameters exhibits a jump at the MI
transition, although LAO shows nothing anomalous around this
range. From Fig.~\ref{fig:nno-v}, we can see that the
discontinuous volume change occurs at about 140 K in 60 nm films
during warming up, in contrast to 200 K in bulk. This is in good
agreement with the T$_C$ obtained by resistivity measurements
(Fig.~\ref{fig:nno-r}). The 200 nm sample behaves similarly, and
its T$_C$ is about 135 K during cooling down. Another important
feature in films is that the volume expansion with respect to the
volume on the metallic side is only 0.053\% for 60 nm sample,
considerably smaller than the bulk value of 0.23\%. The thermal
expansion coefficient $\alpha$ behaves similar in bulk and films.

In bulk NNO, the changes in all three lattice constants contribute
to the overall volume discontinuity, while in epitaxial films,
this is no longer feasible, since the in-plane lattice parameters
are constrained by the underlying substrates. Assuming a volume
change is necessary for MI transition, then in thin films, the
only possible way is to let out-of-plane lattice constant
\textit{c} change accordingly. Therefore, in phase transitions
with volume discontinuity, the substrate clamping effect still
controls the behavior of in-plane lattice parameters, while the
out-of-plane parameters adapt to accommodate the volume change
necessary for the phase transition.

\subsection{SrRuO$_3$}

Bulk SrRuO$_3$ undergoes phase transitions at several
temperatures. Above 950 K (677\textcelsius), SRO is cubic with
Pm$\overline{3}$m symmetry. Between 950 K to 820 K it is
tetragonal I4/mcm. Below 820 K, the unit cell becomes orthorhombic
with space group Pbnm.\cite{Kennedy98} At about 160 K, SRO goes
through another transition to a ferromagnetic phase.

We found domain structure in the SrRuO$_3$ buffer layers at room
temperature, presumably due to tetragonal-orthorhombic transition.
This appears to be a counter-case to the clamping effect we
mentioned before. Notice that we will use orthorhombic notation
for SRO, but still employ the pseudo-cubic notation for LAO. SRO
has a orthorhombic unit cell of about
$\sqrt{2}$\textit{a}$\times\sqrt{2}$\textit{a}$\times$2\textit{a},
where $a$ is pseudo-cubic axis.

The growth temperature 720\textcelsius\ ($\sim$993 K) is
substantially higher than the first phase transition temperature
of SRO. Therefore at growth temperature, SRO is possibly cubic,
with lattice mismatch fully relaxed by forming misfit dislocation.
Upon cooling down to room temperature, it may form poly-domain
structure to relax the strain.\cite{Alpay98,Jiang01} According to
the lattice parameters of pseudo-cubic LAO and orthorhombic SRO,
there are six possible domain orientations.\cite{Jiang98} With
long axis ($c$) of SRO unit cell parallel to the [1 0 0]$_c$ of
LAO lattice, we have a pair of 90 degree twin domains, $A$ and
$A^\prime$. Similarly, with long axis parallel to [0 1 0]$_c$ of
LAO defines $B$ and $B^\prime$ domain pair, and parallel to [0 0
1]$_c$ of LAO defines $C$ and $C^\prime$ domains.

We examined SRO layers in several samples by x-ray reciprocal
space mapping at room temperature. At pseudo-cubic (113)$_c$
position, which would be (206)$_o$ for c domain or (422)$_o$ for
a(b) domains, we observed two SRO peaks splitting along the cubic
[110]$_c$ direction. This arrangement of the two peaks suggests
$c$ domains, with SRO $c$ axis perpendicular to the surface of the
films. The two peaks correspond to (206)$_o$ of one domain and
(026)$_o$ of the other domain from a 90\textdegree\ twin
structure.

\begin{figure}
\includegraphics[scale=1.0]{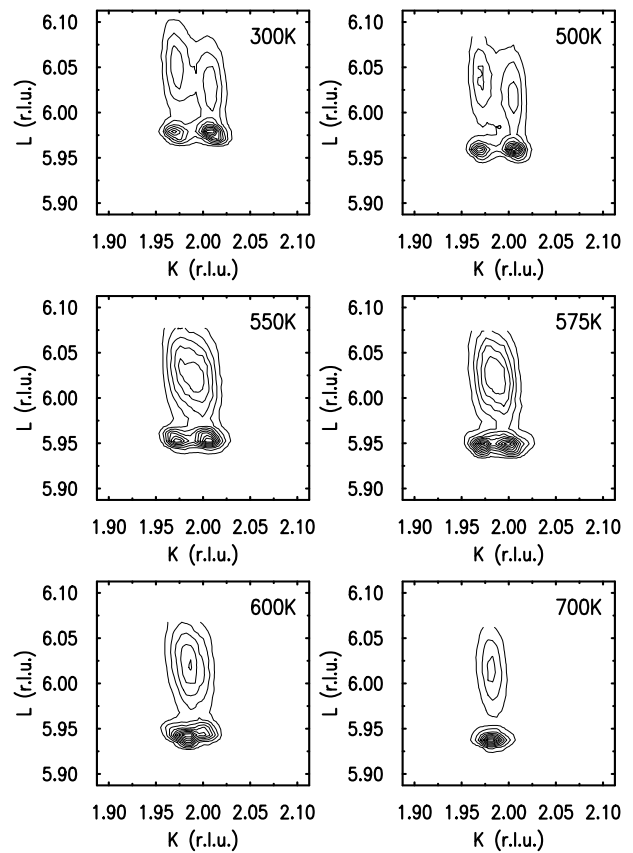}
\caption{\label{fig:sro-map}Reciprocal space mapping shows SRO
forming domain structure. The lower peaks are SRO (2~0~6)$_o$ and
(0~2~6)$_o$. The upper weak peaks are from STO (1~1~3)$_c$. The
unit is in reciprocal lattice unit of SRO.}
\end{figure}

\begin{figure}
\includegraphics[scale=1.0]{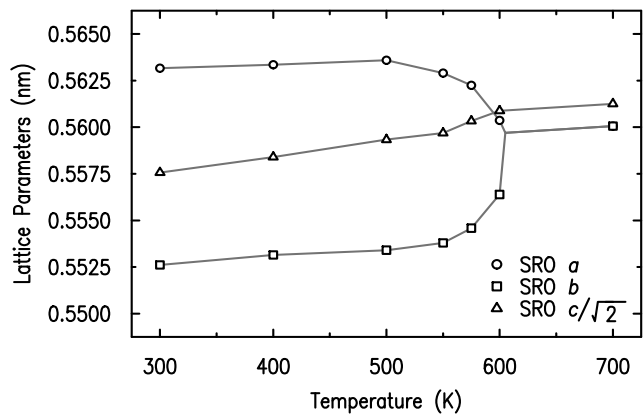}
\caption{\label{fig:sro-lat}Temperature dependence of SRO lattice
parameters.}
\end{figure}

We have performed temperature scans from 300 K to 700 K on a 5 nm
STO / 350 nm SRO film on LAO sample, Fig.~\ref{fig:sro-map}. Both
$hk$ plane and $kl$ plane were scanned. At room temperature, there
are clearly two SRO peaks. From 500 K the peaks began to get
closer, and finally merged at about 600 K. Upon cooling down, the
peak splits at the same temperature, showing a reproducible phase
transition. This T$_C$ is almost 150$\sim$200 K below the bulk
value. The temperature dependence of lattice parameters obtained
from above measurements are shown in Fig.~\ref{fig:sro-lat}.
Overall it is similar to the bulk, though the splitting is bigger.

\section{\label{sec:Discussion}Discussion}

\subsection{Order Parameters}

\begin{figure*}
\includegraphics[scale=1.0]{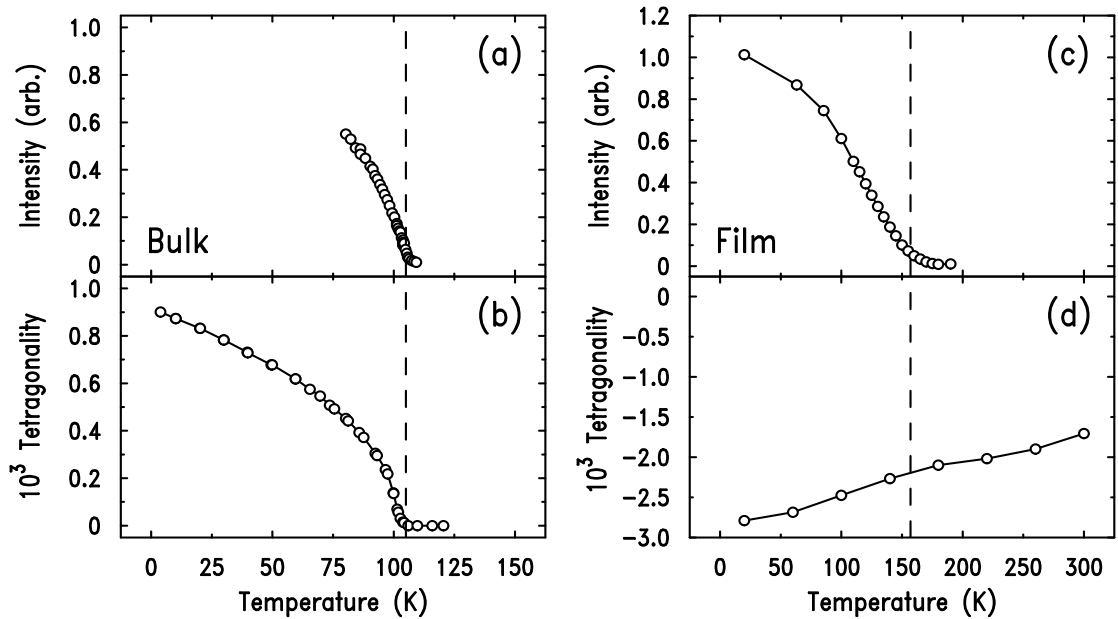}
\caption{\label{fig:orderpar}Order parameters for phase
transitions in STO bulk and films, (a) superlattice peak intensity
in bulk, (b) tetragonality in bulk, (c) superlattice peak
intensity in films, (b) tetragonality in films. Dashed lines
indicate T$_C$. Bulk data from Ref.~\onlinecite{Hayward99,
Hirota95}.}
\end{figure*}

According to Landau theory, the phase transitions of the second
order can be quantitatively described by properly defined order
parameters, which represent the changes in the symmetry in the
crystals when they pass through the phase transition points. The
rotational angle of the TiO$_6$ octahedra, $\varphi$, has been
identified as the order parameter of phase transition in
STO,\cite{Muller68} and denotes the deviation from the perfect
cubic perovskite structure. The intensity of the superlattice peak
is proportional to the square of order parameter.  For bulk STO
samples, above the phase transition temperature T$_C$, the
structure is in cubic symmetry, so there is no superlattice peak.
Below T$_C$, the TiO$_6$ octahedra rotate by an angle $\varphi$,
and superlattice peaks appear. Concurrently, the three lattice
parameters are no longer equal, with c axis longer than the other
two. The changes of the lattice parameters indicates the
tetragonality of the unit cell changes at T$_C$. Therefore, we can
use the tetragonality as a secondary order parameter to describe
this phase transition.\cite{Sato85,Binder01} We define the
tetragonality as:
\begin{equation}
\gamma=\frac{c-\frac{(a_{1}+a_{2})}{2}}{\frac{(a_{1}+a_{2}+c)}{3}}
\end{equation}
where $a_1$, $a_2$ and $c$ are the lattice constants in
pseudo-cubic notation, and the average of two in-plane values
$a_1$ and $a_2$ is used to account for possible experimental
error. For the cubic phase, $\gamma$ is zero. The tetragonal phase
has either positive or negative $\gamma$ depending on if $c$ is
greater than or less than $a$. It is interesting to notice that
the order parameter defined by rotation angle represents a
internal change within the unit cell, while the order parameter
obtained from tetragonality gives the external change in the shape
of the unit cell.

From above results, we see an important change in the nature of
this phase transition in thin films vs bulk. In bulk STO, the
internal and external signals are intimately related, the TiO$_6$
octahedra start to rotate while the shape of the unit cell distort
at T$_C$. Both signatures show sharp transition points at the same
transition temperature of 105 K. In epitaxial thin films, the
external lattice shape is no longer free to change, however, the
TiO$_6$ octahedra still begin to rotate at particular temperature
within this fixed cage. Therefore, although both the order
parameters as defined before are good for bulk, in thin films only
the internal order parameter can clearly indicate the phase
transition. The tetragonality order parameter shows no phase
transition, as illustrated in Fig.~\ref{fig:orderpar}. In other
words, the internal degree of freedom is now decoupled from the
external degree of freedom.

This phenomenon is not unique for the STO system. Similar cases
happen in ferroelectric phase transitions in epitaxial BaTiO$_3$
or other thin films. When bulk BaTiO$_3$ goes through a
ferroelectric phase transition at 120\textcelsius, its lattice
constants change abruptly, while the unit cell volume is
continuous acrossing the T$_C$.\cite{Devonshire49} But in thin
films, there is no obvious change in all three lattice parameters
at the phase transition.\cite{Yoneda93,Terauchi92,Alpay04} However
from the dielectric measurement, Yoneda and coworkers observed
ferroelectricity in the low-temperature phase and paraelectricity
in the high-temperature phase in films. This indicates that there
is a phase transition with respect to the internal displacement of
the Ti and O atoms, although the external structural component of
the phase transition is suppressed in the presence of epitaxial
strain and substrate constraint.

For BTO, the primary order parameter is the spontaneous
polarization $\mathbf{P}$ ($\mathbf{D}$=0), which is proportional
to the internal relative displacement of Ti and O atoms (the
ferroelectricity of the material), and again the tetragonality of
the unit cell is the secondary order parameter.\cite{Cowley76} We
will see exactly the same relationship between the bulk and thin
films as the case of STO. Therefore, from the view point of order
parameters, we have a new kind of phase transition which is not
possible in bulk.

\subsection{Domain Formation}

It is natural to ask why some films show bulk-like relaxation of
lattice parameters at the transitions while others do not. This is
equivalent to asking why some films form domains while others do
not.  We give a qualitative explanation from an energetic point of
view.

If there is a material with a cubic to tetragonal phase
transition, above T$_C$, cubic phase is energetically favored, and
below T$_C$ tetragonal symmetry is favored. During the transition,
some atoms within the material will have slight displacement, from
$X_{film}$ positions in cubic phase to $X_{film}^\prime$ positions
in tetragonal phase. In other words, the potential energy minimums
for these atoms shift, and this is the internal driving force for
the phase transition and domain formation. If external forces are
applied to keep these atoms stay at $X_{film}$ positions even
below T$_C$, the material is strained and the total free energy
increased.

On the other hand, imagine we have a completely flexible epitaxial
film growth on a substrate, the film atoms would always align with
the underlying substrate atoms. The substrate creates a potential
energy minimum at each of the $X_{sub}$ positions directly above
the substrate atoms, and these minima only move with substrate. If
we need to move some atoms away from these $X_{sub}$ positions, we
must supply some energy. And this again raises the total free
energy of the system.

Combining these two pictures, we can see there are two competing
energies during this process, involving internal strain energy
from film and interface strain energy due to substrate. For a
thick film, at the growth temperature, the two minimum $X_{film}$
and $X_{sub}$ are aligned with each other through formation of
misfit dislocations. When the temperature is below T$_C$, the film
has the tendency to complete the structural phase transition, and
this makes the two potential energy minima separated into
$X_{film}^\prime$ and $X_{sub}$. The deviation of an atom from one
minimum will raise the corresponding strain energy, and we expect
the increase to be small if the displacement is small.

Generally the structure change is minor when temperature is near
T$_C$, so just below T$_C$ the potential minimum at
$X_{film}^\prime$ due to film is likely to be close to $X_{sub}$
position and possibly shallow than $X_{sub}$. Therefore the
driving force for the domain formation is small and the film atoms
will still follow substrate. No domain forms. When temperature is
considerably lower than T$_C$ and the depth of minimum at
$X_{film}^\prime$ exceeds that at $X_{sub}$, the stable position
moves to $X_{film}^\prime$, therefore forming domain is
energetically favored. However there will be a barrier between the
two minima, and in order to overcome the barrier to reach
$X_{film}^\prime$, the atoms need some thermal energy.

For STO, there can be two scenarios, (i) since the structural
change is subtle, the potential minimum at $X_{film}^\prime$ may
always be shallower than that at $X_{sub}$; (ii) since the T$_C$
is far below the growth temperature, even if the film has a
tendency to form domain below T$_C$, the thermal energy of the
film atoms is too low to overcome the barrier to the new position.
So even though the minimum at $X_{film}^\prime$ is lower, the
atoms still stay at the minimum $X_{sub}$ defined by the
substrate. In both cases, it appears macroscopically that the film
is frozen to the substrate, without any obvious in-plane lattice
change. The domain formation is suppressed by the substrate
constraint. BaTiO$_3$ and NdNiO$_3$ films fall into exactly the
same situations.

For SRO, the phase transition occurs at above 820 K in bulk. If we
assume the film has the tendency of a phase transition at a
similar temperature, then the atoms still have substantial thermal
energy, or mobility, to move around. If at a certain temperature
the move from $X_{sub}$ to $X_{film}^\prime$ is energetically
favorable, the film atoms will move to the minimum at
$X_{film}^\prime$ defined by the internal strain energy, then new
domains form. Further experiments are on going to confirm this
explanation.

\subsection{Lattice Parameters vs Thickness}

\begin{figure}
\includegraphics[scale=1.0]{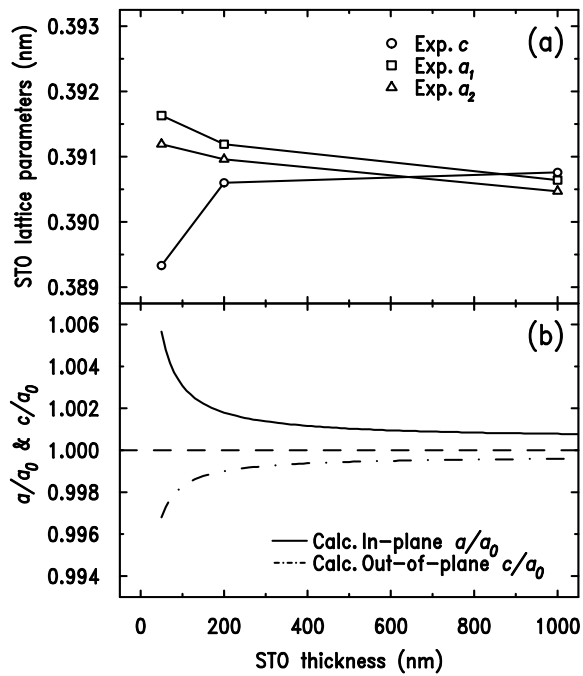}
\caption{\label{fig:lat-sto}Lattice parameters for STO films of
different thickness on 350 nm SRO layers. Both experiment data (a)
and calculated value (b) are shown. From Ref.~\onlinecite{Ban04}}
\end{figure}

\begin{figure}
\includegraphics[scale=1.0]{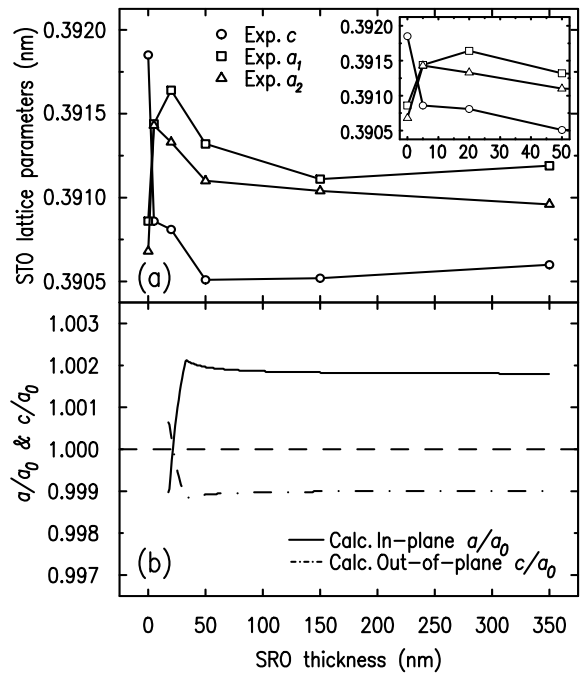}
\caption{\label{fig:lat-sro}Lattice parameters for 200 nm STO
films on SRO buffer layers of varying thickness. (a) experiment
results, inset is the detail of thin SRO region, (b) calculated
STO lattice parameters. From Ref.~\onlinecite{Ban04}}
\end{figure}

From XRD measurements of lattice constants, we can see that both
film thickness and buffer layer thickness influence the strain
state of the STO films. Fig.~\ref{fig:lat-sto} and
Fig.~\ref{fig:lat-sro} show both the out-of-plane and in-plane
lattice parameters at room temperature for different film / buffer
combinations. For films grown on the same substrates and buffer
layers, when STO film is very thin, the misfit strain is not fully
relaxed, so the in-plane lattice constants are expanded, while the
out-of-plane lattice constant is reduced in order to keep roughly
the same unit cell volume. As the STO thickness increases, the
strain within the STO films becomes more and more relaxed, so both
the out-of-plane and in-plane lattice parameters move towards the
equilibrium bulk values. The thickest film, the 1000 nm STO / 350
nm SRO sample, is already indistinguishable from cubic.
Considering the experimental error, its three lattice parameters
are almost the same. Also it behaves more like bulk crystals, with
T$_C$ of 122K, most close to 105K.

For the same STO films grown on different SRO layers, the
variation of STO lattice constants becomes more complicated. Since
the lattice constants of SRO are larger than that of STO, so it
seems reasonable to expect thick SRO layers give rise to a larger
strain in STO films. But the results turn out to be just opposite,
the thinnest SRO layer creates biggest deformation in the STO
films. Increasing the SRO thickness, the in-plane lattice
parameters in STO films keep decreasing. The out-of-plane lattice
constant c is also decreasing, and always smaller than in-plane
ones. An interesting phenomenon is all three parameters are larger
than the STO bulk values, which means the volume of the unit cell
is bigger than the bulk.

As seen from Fig.\ref{fig:lat-sro}, the STO lattice constants
change drastically when SRO thickness goes from 0 nm to about 20
nm. If STO is grown directly on LAO substrate, the two in-plane
lattice parameter $a_1$ and $a_2$ are smaller than out-of-plane
$c$, however, they are still larger than the bulk values. That is
to say, STO is expanded by LAO, which is contradictory to the
common thought. This indicates that the strain states in the films
also rely on other factors, such as the differences between the
thermal expansion coefficients of films and substrates, and
defects in the films.

To understand the experiment results, we have calculated the
dependency of the STO lattice parameters upon the thickness of
both STO and SRO layers. The details were published
elsewhere.\cite{Ban04} The theoretical model employed takes into
account the stress relaxation due to formation of orthorhombic
polydomain structure in the SRO buffer layer, as well as the
formation of misfit dislocations at the LAO / SRO and STO / SRO
interfaces. It has been shown that the internal stress level in
films can be controlled using buffer layer that exhibits a
structural phase transition.

Taking into account the multiple strain relaxation mechanisms, the
in-plane and out-of-plane lattice parameters of STO films as a
function of both STO and SRO layer thickness can be calculated, as
shown in Fig.~\ref{fig:lat-sto} and Fig.~\ref{fig:lat-sro}. In
general the calculations is in good agreement with experiment
data.

\subsection{Phase Diagram}

It is well known that T$_C$ in films change with different film
thickness or different substrates. Generally, one would like to
parameterize the control by a simple coordinate and construct an
appropriate phase diagram. The most commonly discussed parameter
for films is strain. Another candidate for the STO phase
transition might be tetragonality at room temperature. Thus we
would like to verify a Temperature-Strain phase diagram and a
Temperature-Tetragonality phase diagram.

The phase transitions in strained SrTiO$_3$ epitaxial films have
been theoretically calculated and discussed by Pertsev et
al.\cite{Pertsev00} The misfit strain-temperature phase diagram
they predicted shows that, except for the high-temperature
tetragonal phase (HT) being the distorted prototypic cubic phase,
there are two purely structural tetragonal and orthorhombic states
(ST and SO) at lower temperature, and the phase transition
temperature is strongly related to the misfit strain. The minimum
of T$_C$ is at 105K with zero misfit strain, while both positive
strain and negative strain increase T$_C$.

\begin{figure}
\includegraphics[scale=1.0]{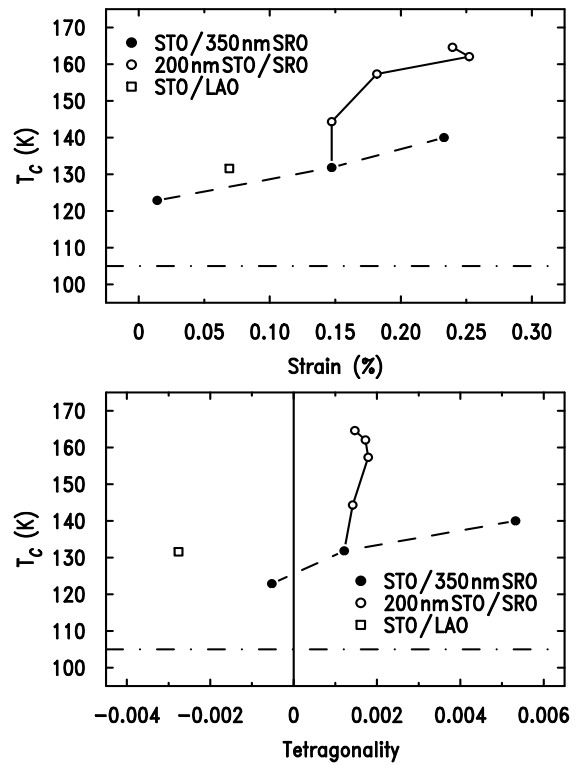}
\caption{\label{fig:tc-strain}Dependence of T$_C$ on (a) in-plane
strain, (b) tetragonality}
\end{figure}

Since there is no indication of the structural phase transitions
with respect to the changes in lattice constants, we obtained the
phase transition temperatures by monitoring the superlattice
peaks, which corresponds to the rotation of the TiO$_6$ octahedra.
In Fig.~\ref{fig:tc-strain}(a), we plot the relationship between
T$_C$ of STO films vs in-plane strain. The general rule for effect
of in-plane strain on T$_C$ is that, the larger the in-plane
strain, the higher T$_C$. The strain in our samples ranges from
0.01\% to about 0.25\%. In this range, the trend of our
measurements are in general agreement with the calculations. The
phase transition temperatures increase with increasing in-plane
strain. Above the transition temperature, the STO film is a
high-symmetry tetragonal phase due to the biaxial strain and
substrate constraint. However, there are some differences. They
predicted that T$_C$ would increase very slowly for tensile
strains, not higher than 115 K for 2\% strain. But our results
show remarkably larger increases in T$_C$ even for much smaller
strains, about 160 K for 0.25\% strain. In addition, it is
possible to induce the same amount of in-plane strain in STO films
by varying either STO thickness or SRO layer thickness. However
the transition temperatures are different for these two cases,
thus T$_C$ vs strain is not a simple monotonic function.

From Fig.~\ref{fig:tc-strain}(b) T$_C$ vs tetragonality of STO
unit cell, we also found there is no simple relationship between
them. The main factor influencing the tetragonality is the
thickness of STO itself. The thin STO films tend to have high
tetragonality while thick STO films is more bulk-like with
tetragonality close to 0. With the same STO thickness of 200 nm,
the films on different SRO layers have almost the same
tetragonality. However the T$_C$s of these films are quite
different. That is to say, tetragonality is not an adequate factor
to dictate T$_C$. This is also consistent with fact that the
tetragonality of each film does not change across the phase
transition.

Some reasons for the discrepancy between theory and experimental
data may include: (i) an ideal cubic substrate was used for
calculations, but the real buffer-substrate systems usually have
mosaic structures and domains. This leads to inhomogeneous strain
conditions across the films; (ii) the different definition of
T$_C$ we used may give a slightly higher T$_C$. However, we
believe the primary reason is because film systems are inherently
more complicated than the model proposed. The strain state in a
film is due to more than the difference between the lattice
parameters of film and substrate. On one hand, two strain
relaxation mechanisms, forming misfit dislocations at the
film-substrate interface and forming domain structures within the
films, greatly reduce the actual misfit strain comparing with that
calculated by respective bulk values. On the other hand, the
difference between the thermal expansion coefficients of films and
substrates, together with substrate constraint, create another
kind of strain, which depends on temperature, but not on the
lattice constants of films and substrates. Furthermore,
non-equilibrium defects increase unit cell volume,\cite{Tarsa96,
Peng03} and thus change strain in a manner different than
epitaxial mismatch.

Adding together all these complications, the T$_C$ vs strain is
not a simple function in real films, and an equilibrium phase
diagram may not be generally applicable.

\section{Conclusion}

Through a systematic study of three distinctive film systems, we
have a more comprehensive understanding on the phase transitions
in epitaxial perovskite films. The final strain state of film, and
thus the T$_C$, results from competition between multiple
mechanisms, and may not be ascribed to a single parameter, such as
misfit strain. There are a few general rules, such as thickness
dependence of strain and substrate constraint, which are followed
by all film systems. However, a particular film system may possess
its own unique features, which must be characterized individually.

We have studied the structural phase transitions in strained STO
films grown on LAO substrates with SRO buffer layers. As indicated
by the superlattice peaks associated with the lower temperature
tetragonal phase, the phase transitions occur at higher
temperature in strained films, with larger in-plane strain giving
higher T$_C$. However, there is no obvious indication of this
transition from temperature dependence of the lattice parameters,
and tetragonality is no longer a good secondary order parameter.
This is an important change in the nature of this phase
transitions. Due to the epitaxial strain and substrate clamping
effect, the internal degree of freedom, the rotation of the
TiO$_6$ octahedra, decouples from the external degree of freedom.
Our data show that both the thickness of STO thin films and SRO
buffer layers influence the magnitude of the strain within STO
films. The in-plane strain decreases with increasing STO
thickness, however, thinner SRO layers give rise to larger
in-plane strains in STO films. This result is in good agreement
with our calculations. Combining the T$_C$ and strain
measurements, it is found that the in-plane strains have
considerably larger effect on T$_C$ than previously predicted.

From our results of NdNiO$_3$ films grown on LAO, it is much
clearer that substrate constraint is an important parameter for
epitaxial films. The in-plane lattice parameters of film are tied
down by the underlying substrate, following the trend of substrate
through the whole temperature range, without change even across
the phase transition in films. This leaves the out-of-plane
lattice parameter the only variable to accommodate the volume
change during the phase transition. This substrate clamping effect
has a profound influence on the strain within film, and can induce
a substantial change in phase transition temperature T$_C$.

SrRuO$_3$ films show us even more complications. In this system we
observed domain formation at elevated temperature. This means the
substrate constraint sometimes may give way to other much stronger
strain relaxation mechanism. This phenomenon illustrates that the
final strain states in films are the results of interaction
between multiple competing mechanisms, such as internal strain
energy within films and interface strain energy.

\begin{acknowledgments}
BW thanks the Cottrell Scholar Program of the research corporation
for partial support of this work. This material is based upon work
supported by the National Science Foundation under Grant No.
DMR-0239667 (BW, FH), DMR-0132918 (ZB, PA), and DMR-9702632 (XX).
Work at Brookhaven is supported by Division of Material Sciences,
U.S. Department of Energy under contract DE-AC02-98CH10886. The
work at UConn was also supported by the University of Connecticut
Research Foundation.
\end{acknowledgments}

\bibliography{sto}

\end{document}